\documentclass{INTERSPEECH2023}
\pdfoutput=1

\interspeechcameraready

\title{SememeASR: Boosting Performance of End-to-End Speech Recognition against Domain and Long-Tailed Data Shift with Sememe Semantic Knowledge}
\name{Jiaxu Zhu$^{1}$, Changhe Song$^{1,2,\dagger}$\thanks{$\dagger$ Equal contribution.  * Corresponding author.}, Zhiyong Wu$^{1,2,3,*}$, Helen Meng$^3$}
\address{
  $^1$Shenzhen International Graduate School, Tsinghua University, Shenzhen, China\\
  $^2$Peng Cheng Lab, Shenzhen, China \\
  $^3$The Chinese University of Hong Kong, Hong Kong SAR, China}
\email{\{zhu-jx21, sch19\}@mails.tsinghua.edu.cn, zywu@sz.tsinghua.edu.cn, hmmeng@se.cuhk.edu.hk}

\begin{document}

\maketitle 
\begin{abstract}
Recently, excellent progress has been made in speech recognition. However, pure data-driven approaches have struggled to solve the problem in domain-mismatch and long-tailed data. Considering that knowledge-driven approaches can help data-driven approaches alleviate their flaws, we introduce sememe-based semantic knowledge information to speech recognition (SememeASR). Sememe, according to the linguistic definition, is the minimum semantic unit in a language and is able to represent the implicit semantic information behind each word very well. Our experiments show that the introduction of sememe information can improve the effectiveness of speech recognition. In addition, our further experiments show that sememe knowledge can improve the model's recognition of long-tailed data and enhance the model's domain generalization ability.
\end{abstract}

\section{Introduction}
Automatic Speech Recognition (ASR) is a technology that converts audio into text. 
In recent years, end-to-end (E2E) ASR has attracted a lot of attention and has made great progress. E2E ASR can convert audio to text using a single network model, greatly simplifying the training and inference process. There are three main types of E2E ASR models: connectionist temporal classification (CTC) \cite{graves2006connectionist}, recurrent neural network transducer (RNN-T) \cite{graves2012sequence, wang2019exploring}, and attention based encoder-decoder (AED) \cite{chorowski2014end, chan2016listen, luo2021simplified}. 
E2E ASR models achieve excellent results by leveraging large amounts of training data, which is the so-called pure data-driven approach.

However, pure data-driven approaches suffer poor recognition of long-tailed data and poor domain generalization 
due to the performance depends entirely on the training data, even though they are extremely characteristic and unevenly distributed.
\begin{figure}[t]
	\centering
	\includegraphics[width=\linewidth,scale=1.00]{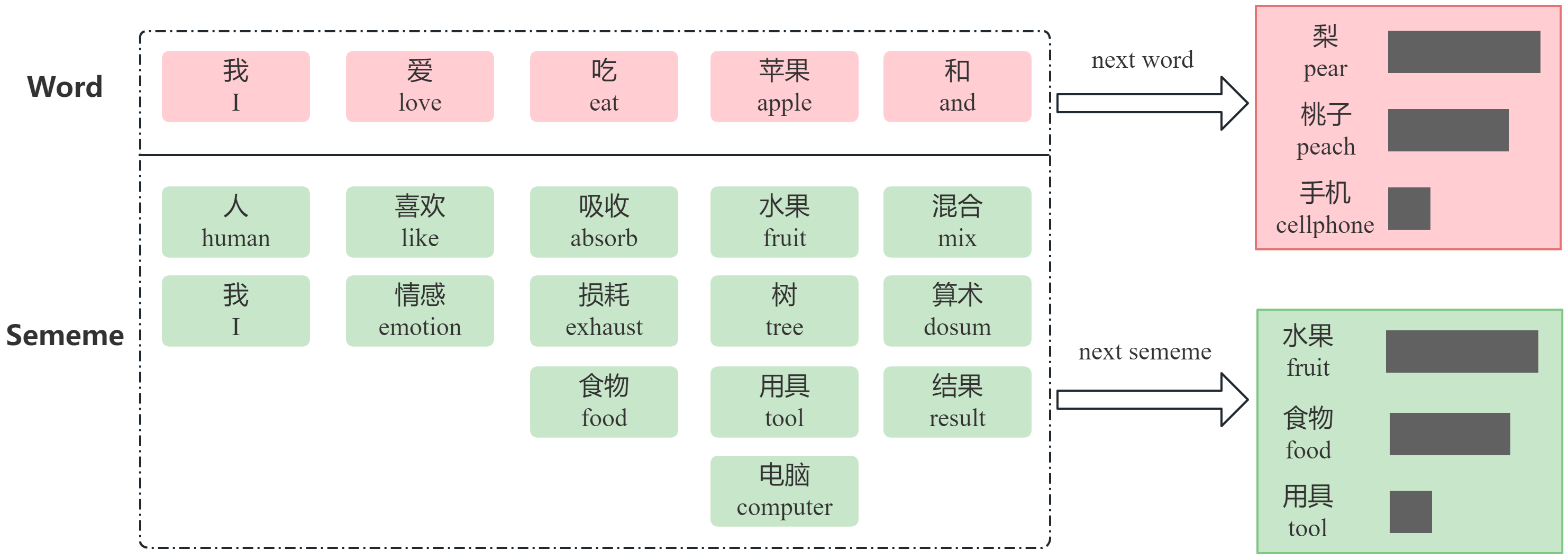}
	\caption{Examples of words and sememes. The red blocks are the words, and the green blocks are the set of sememes corresponding to the words above. The next word can be predicted more accurately by combining sememe information.}	
	\label{sememe_ex}	
\end{figure}
As poor recognition of long-tailed data and weak domain generalization are brought about by the training data itself, 
introducing external knowledge information can help alleviate this problem.
The semantic knowledge information implied behind textual data has become a hot topic of research.
As shown in Figure \ref{sememe_ex}, sememe is defined as the minimum semantic unit of languages in linguistics.
Sememe knowledge has been widely studied in the field of natural language processing (NLP) \cite{niu2017improved, gu2018language, qi2021sememe, zhang2020enhancing}.
Compared to rough text data, sememe knowledge is more accurate and fundamental and has been refined by professional scholars over many years.
The sememe  knowledge, capable of representing the semantic information of any word, is more stable and robust and is not affected by the data.
A knowledge-driven approach based on sememe knowledge can alleviate the problems of data-driven approaches and improve the effectiveness of models in long-tailed data problems.

As far as we know, we are the first to introduce sememe into ASR.
In this paper, we propose a sememe-based semantically enhanced ASR model called SememeASR, which improves the semantic capability of the model by improving the semantic information of the text representation and adding sememe prediction tasks. 
Inspired by \cite{niu2017improved, zhang2020enhancing}, while the traditional data-driven approach is not able to fully exploit the rich semantic information of text,
adding the information of the sememe can enrich the semantic representation of text.
Our experiments have shown that different methods of adding sememe information to the model can improve the recognition ability and enhances the model's ability to recognize long-tailed data and somewhat enhances the model's domain generalization capability.


\section{Methodology}
In this section, we review the architecture of the baseline hybrid CTC/AED model in Section 2.1. Then our proposed method will be described in Section 2.2, 
which aims to apply sememe-based semantic knowledge to improve the ability of the ASR model, thereby improving the recognition ability of long tail data and enhancing the domain generalization ability.
\subsection{Hybrid CTC/AED ASR Model}
\begin{figure}[h]
	\centering
	\includegraphics[width=0.8\linewidth,scale=1.00]{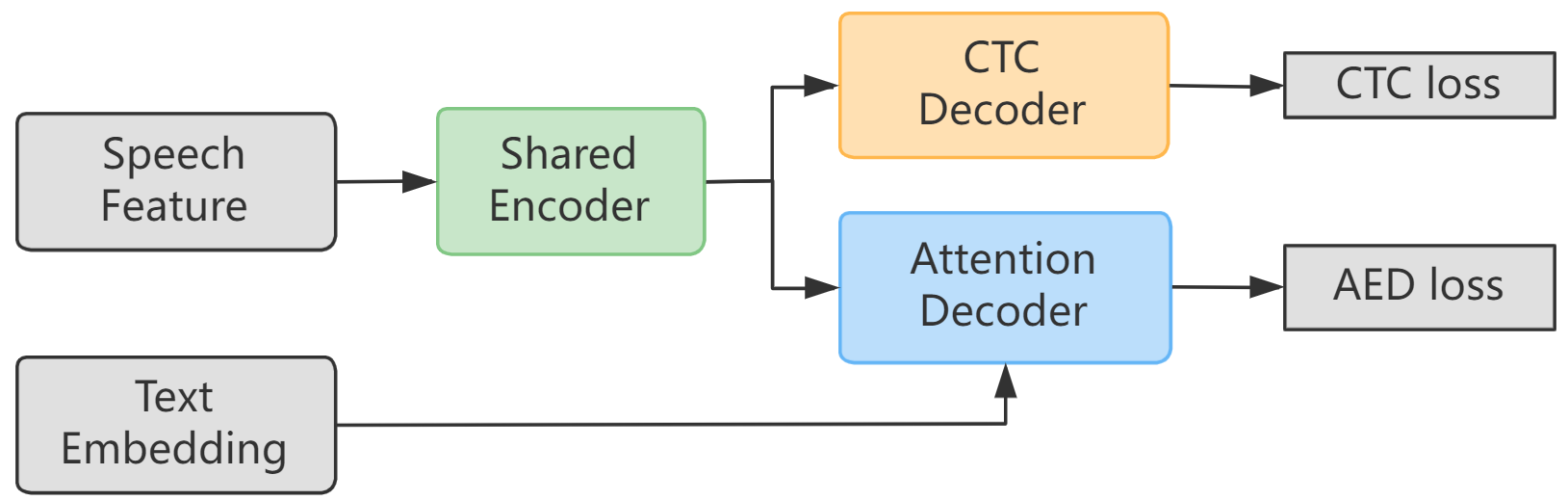}
	\caption{Architecture of Hybrid CTC/AED ASR  model.}	
	\label{BaselineASR}
 \vspace{-0.5cm}
\end{figure}

The baseline E2E ASR model we choose in our experiment is similar to the one presented in WeNet \cite{yao21_interspeech}, which uses both CTC and Attention-based Encoder-Decoder (AED) loss during training to speed convergence and is also a relatively good one among a series of state-of-the-art approaches \cite{peng2022branchformer,ren2022improving}.
As depicted in Figure \ref{BaselineASR}, the hybrid CTC/AED ASR model mainly contains three parts, a \textit{Shared Encoder}, a \textit{CTC Decoder}, and an \textit{Attention Decoder}.
The \textit{Shared Encoder} consists of a convolution subsampling layer containing two convolutional layers with stride 2 for downsampling, a linear projection layer, and a positional encoding layer, followed by multiple Conformer \cite{gulati20_interspeech} encoder layers. 
The \textit{CTC Decoder} consists of a linear layer and a log softmax layer. 
The \textit{Attention Decoder} consists of a positional encoding layer, multiple Transformer \cite{VaswaniSPUJGKP17} decoder layers, and a linear projection layer.

\subsection{Proposed Model Architecture}
The main idea of our paper is to introduce sememe-based semantic information into ASR simply and effectively. Inspired by \cite{zhang2020enhancing}, we use three simple but effective strategies:
1) adding sememe prediction auxiliary task with sememe loss in a multi-task learning manner,
2) simply adding sememe information to the textual representation, or
3) employing \textit{Sememe Encoder} to improve the semantic representation ability of the text.

\subsubsection{Sememe Prediction Task}
\begin{figure}[h]
	\centering
	\includegraphics[width=0.95\linewidth,scale=1.00]{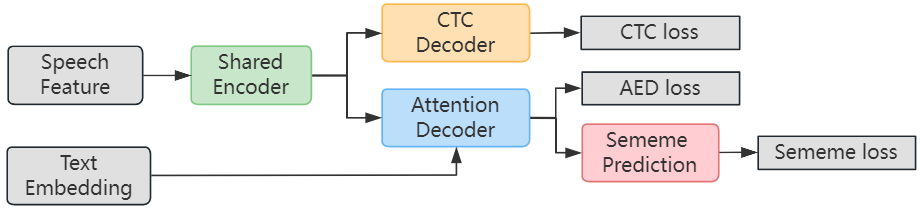}
	\caption{Architecture of SememeASR-SP.}	
	\label{SP}	
 \vspace{-0.5cm}
\end{figure}

As depicted in Figure \ref{SP}, we add the \textit{Sememe Prediction} task with sememe loss after the \textit{Attention Decoder}.
A multilabel classification task can be used to build the sememe prediction task, which seeks to predict sememes for the following token.
Understanding semantics is closely related to predicting the following token's sememes, which is frequently easier to learn than directly modeling the likelihood of the next token.
Given current contextualized representation $\mathbf{g}$ from Transformer in \textit{Attention Decoder}, we estimate the probability of sememe $s$ associated with next token $t$ as showed in Equation \ref{SememeASR-SP}:
\begin{equation}
    p(t, s)=\sigma\left(W \mathbf{g}+b\right)
    \label{SememeASR-SP}
\end{equation}
where $W$ and $b$ are the weight and bias associated with sememe $s$, $\sigma$ is the sigmoid activation function. 
We have named the model that uses this approach  \textbf{SememeASR-SP}.

\subsubsection{Semantically Enhanced Text Representation}
\begin{figure}[h]
	\centering
	\includegraphics[width=0.9\linewidth,scale=1.00]{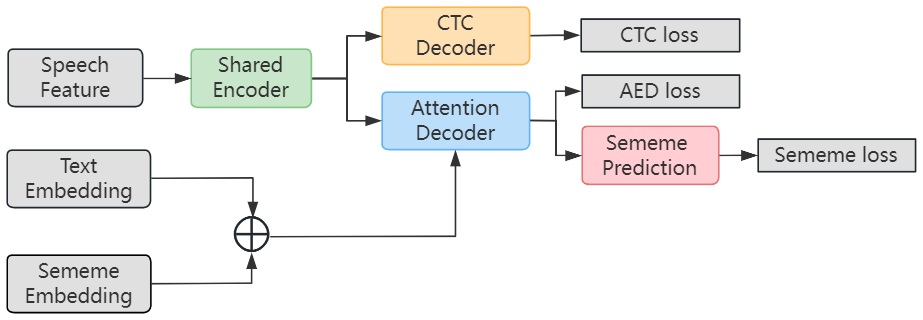}
	\caption{Architecture of SememeASR-SE.}	
	\label{SE}	
 \vspace{-0.5cm}
\end{figure}

Traditional data-driven training methods in ASR have difficulty in mining the rich semantic information in the text.
Some approaches use BERT \cite{bai2021fast} to provide semantic information, which helps to increase the semantic information of ASR, but this semantic information is still relatively shallow and poorly interpretable.
The introduction of sememe-based semantic information not only provides rich semantic information but also has strong interpretability.
As depicted in Figure \ref{SE}, one simple method to add the sememe information is 
averaging the corresponding sememe of each token to get the sememe embedding corresponding to the text, and then adding it to the text embedding. 
We denote original text embedding as $E=(e_1, e_2, ...,e_i,...,e_I)$, sememe embedding as $C=(c_1, c_2, ..., c_i, ..., c_I)$ and final semantically enhanced text embedding as $\hat{E}=(\hat{e_1}, \hat{e_2}, ...,\hat{e_i},...,\hat{e_I})$, where $I$ is the number of tokens in the text sequence.
Formally, we have:
\begin{equation}
    c_i=\frac{1}{n_{t}} \sum_{s \in S(t)} {q}_{s}
    \label{sememe_emb}
\end{equation}
\begin{equation}
\hat{{e}}_{i}={c}_{i}+{e}_{i}
    \label{sememe_se}
\end{equation}
where $t$ represents the $i$-th token of text sequence, $S(t)$ refers to the sememe set associated with token $t$,  $n_t$ is the number of sememe entries of token $t$, ${q}_{s}$ refers to the embedding of the sememe $s$. And ${c}_{i}$ is formed by averaging the corresponding embeddings of all sememes of token $t$ as shown in Equation \ref{sememe_emb}. The sememe enhanced token embedding $\hat{{e}}_{i}$ is thus derived by adding ${c}_{i}$ and ${e}_{i}$ as showed in Equation \ref{sememe_se}. Then the semantically enhanced text representation $\hat{E}$ is directly fed into the \textit{Attention Decoder}.
We have named the model that uses this approach \textbf{SememeASR-SE}.

\subsubsection{Sememe Encoder}
\begin{figure}[h]
	\centering
	\includegraphics[width=0.9\linewidth,scale=1.00]{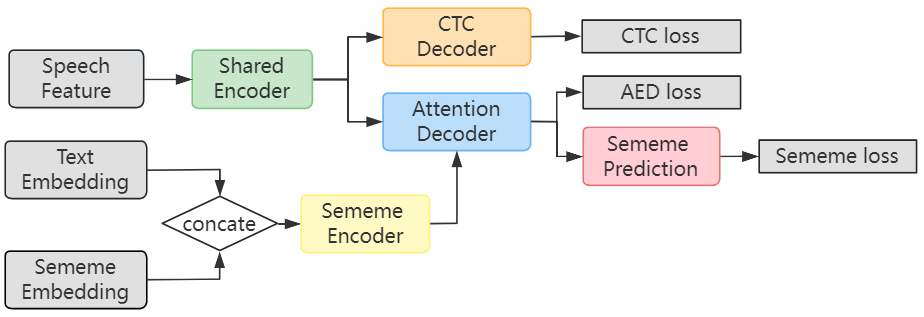}
	\caption{Architecture of SememeASR-SEP.}	
	\label{SEP}	
 \vspace{-0.2cm}
\end{figure}

To make full use of semantic information, we explore better ways to incorporate 
sememe information and improve text representation.
Inspired by \cite{newell2016stacked}, a bottom-up and top-down network structure is used to compose the image from high to low resolution to extract stronger semantic features, and the top-down process then raises the resolution to enhance the designated features. 
Similarly, we use stacked linear layers to achieve a similar effect of the bottom-up and top-down network structure, which we refer to as \textit{Sememe Encoder} as shown in Figure \ref{SEP}. 

\begin{figure}[h]
\center{\includegraphics[width=0.85\linewidth]  {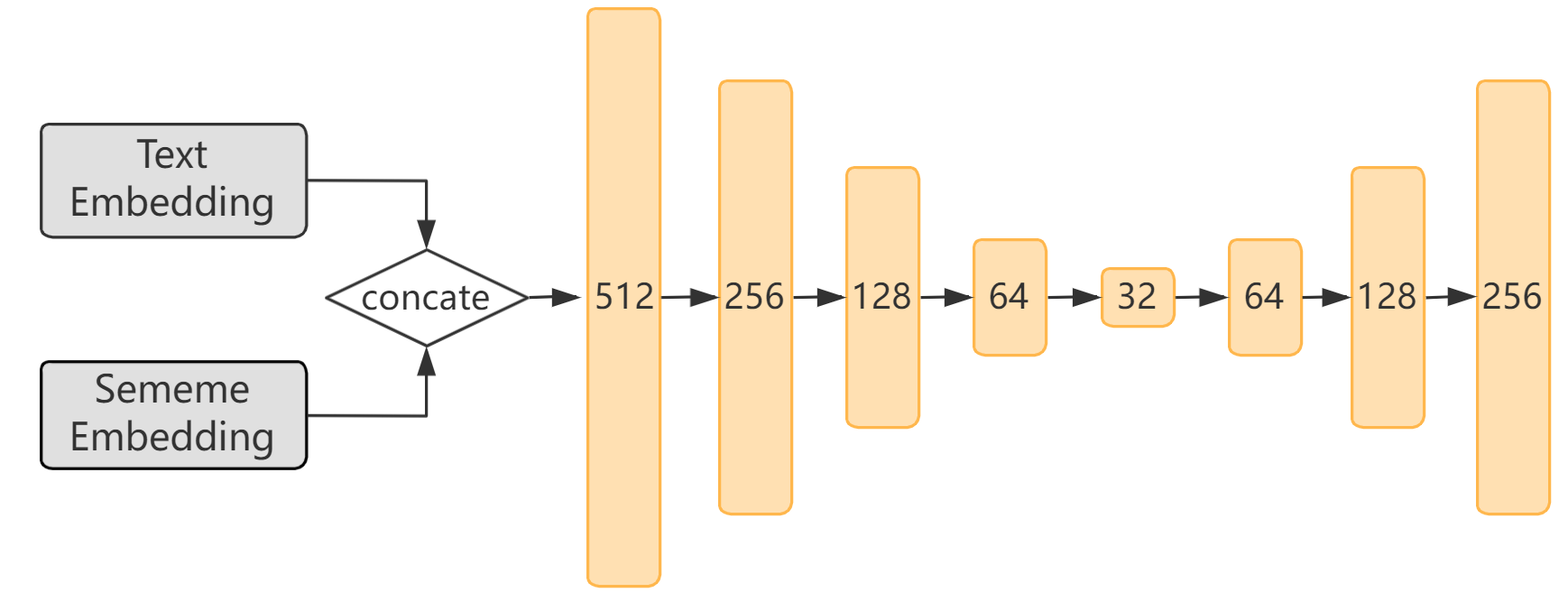}} 
\caption{\label{sememe_encoder} Illustration of the dimension change in Sememe Encoder that consists of multiple linear layers.} 
\vspace{-0.5cm}
\end{figure}

We reduce the dimension of the concatenated text and sememe representation to extract stronger semantic features, and then increase the dimension to strengthen the formed features.
The dimension changes are shown in Figure \ref{sememe_encoder}. In our experiment, the dimension of text embedding and sememe embedding is 256. Dimension after embedding concatenation will be 512 and the \textit{Sememe Encoder} will transform the dimension to 256 which matches the dimension of \textit{Attention Decoder}.
We name the model assigned to this method as \textbf{SememeASR-SEP}.

\subsubsection{SememeASR Loss Function}
Just like the WeNet \cite{yao21_interspeech}, we also combine the CTC and AED losses during training to speed convergence.
Furthermore, we calculate the binary cross-entropy loss of sememe prediction as sememe loss $L_{SE}$ to enhance the model's modeling of sememe-based semantic information.
Equation \ref{loss} defines the SememeASR objective, $L_{C T C}, L_{A E D}$ are the CTC and AED losses respectively, $\lambda \in(0,1)$ and  $\alpha\in(0,1)$ are hyper-parameters where $\lambda$ balances the importance of CTC and AED loss while $\alpha$ balances the importance of AED loss and Sememe loss. 
Finally, the training loss can be represented as follows:
\begin{equation}
Loss=\lambda L_{C T C}+(1-\lambda) [\alpha L_{A E D}+ \\ (1-\alpha) L_{SE}]
\label{loss}
\end{equation}

\section{Experiments}
\subsection{Datasets}
In this paper, we train our proposed E2E ASR on public Mandarin datasets Aishell-1 \cite{bu2017aishell}.
The Aishell-1 corpus consists of 178 hours of labeled speech collected from 400 speakers. The content of the datasets covers 5 domains including Finance, Science and Technology, Sports, Entertainments, and News. To compare the domain adaptation ability
of ASR in the text domain while minimizing the influence of differences in the acoustic environment, we 
test the trained model on
another public Mandarin dataset Aishell-2 \cite{du2018aishell} that has a similar acoustic environment for sound recording but the corresponding text contents cover different text domains.
The Aishell-2 corpus consists of 1000 hours of labeled speech collected from 1991 speakers. 
The content of Aishell-2 corresponds to the domains of voice commands, digital sequence, places of interest, entertainment, finance, technology, sports, English spellings, and free speaking without specific topics.
To further illustrate the robustness of our method, we also 
conduct evaluation experiment
on different
domains 
with WenetSpeech \cite{zhang2022wenetspeech}, which is a multi-domain Mandarin corpus consisting of high-quality labeled speech but a relatively more complex acoustic environment than Aishell-1.

\subsection{Experimental Setup}
For all experiments, we use the open-source WeNet toolkit \cite{yao21_interspeech} to build both the hybrid CTC/attention baseline and our proposed SememeASR. And we used the default values in the WeNet for the main parameters  which have been validated by the WeNet contributor.
The input features are 80-dimensional log Mel-filterbank (FBank) computed on 25ms window with 10ms shift.
We use SpecAugment \cite{park19e_interspeech} and speed perturb for data augmentation.
We choose 4233 characters (including 〈blank〉, 〈unk〉, 〈sos/eos〉 labels) as model units for Aishell-1.

We construct the foundation model using 12 Conformer blocks in the \textit{Shared Encoder} and 6 Transformer blocks in the \textit{Attention Decoder}.
We employ $h$ = 4 parallel attention heads in both Conformer block and Transformer block.
For every layer, we use $d_k = d_v = d_{model}/h$ = 64, $d_{ffn}$ = 2048. 
Our proposed SememeASR model adds sememe encoder module and sememe prediction auxiliary task based on the baseline.

We train the model with Adam Optimizer \cite{VaswaniSPUJGKP17} for at most 240 epochs with a batch
size of 12.
And \textit{learning rate} = 0.002, \textit{warm up} = 25000, and gradient clipping at 5.0. 
Additionally, during training, we employ the gradient accumulation method, in which the gradients are modified every four batches.
Moreover, we employ label smoothing of value $\epsilon_{l s}$ = 0.1 and dropout rate of $P_{drop}$ = 0.1. 
We set the weight $\lambda$ of the CTC branch during joint training to 0.3. Considering that $\alpha$=0.3 achieves better results in our experiment, we choose it as the weight parameter of sememe loss. 
During joint decoding, we set the CTC-weight $\lambda$ to 0.5.
To avoid overfitting, we averaged the 30 best model parameters in the development dataset. 

\section{Experimental results}
The performance of the models is evaluated based on character error rates (CER) without external language models. 
Our experimental results are mainly based on the attention-rescore two-step decoding method.

\subsection{Results of Different Dataset}
We first present the results on the Aishell-1 test dataset.
Table \ref{cer} compares the CER results of different models.
From the results of Aishell-1, we can see our proposed SememeASR model is better than the baseline hybrid CTC/AED model.
\begin{table}[h]
  \centering
  \caption{Comparison of CER on Aishell-1 and Aishell-2}
    \begin{tabular}{cc|cc}
    \hline
    & Aishell-1 &  Aishell-2 & Aishell-2\\
    \textbf{Model} & \textbf{test} & \textbf{dev} & \textbf{test}  \\
    \hline  Baseline (CTC/AED)  & 4.56 & 12.18 & 12.03 \\
    \hline SememeASR-SP  & 4.53 & 11.99 & 12.16 \\
    \hline SememeASR-SE & 4.59 & 11.98 & \textbf{11.95} \\
    \hline SememeASR-SEP & \textbf{4.53} & \textbf{11.93} & 12.03\\
    \hline
    \end{tabular}
  \label{cer}
  \vspace{-0.3cm}
\end{table}

We also compare the results on the Aishell-2 test and dev datasets, which have a similar acoustic environment with Aishell-1 but cover different text domains.
From the results of Aishell-2 in Table \ref{cer}, we can see that our proposed SememeASR model outperforms the baseline on the new domain data. 
This indicates the improvement in the domain generalization capability of our proposed model.

\begin{table}[htbp]
  \centering
  \caption{Performance on different domains of WenetSpeech}
    \begin{tabular}{c|c|ccc}
    \hline
    & & & \textbf{SememeASR}\\
    \textbf{Domain} & \textbf{Baseline} & \textbf{SP} & \textbf{SE} & \textbf{SEP}\\
    \hline
    audiobook  &   \textbf{15.29}  & 15.47  & 15.59  & 15.69\\
    commentary  &   37.74  & 37.09  & 37.00  & \textbf{36.52} \\
    documentary  &   41.04  & 40.53  & \textbf{40.24}   & 40.36\\
    drama  &   56.24  & \textbf{54.39}  & 55.68  & 55.10\\
    interview  &   38.44  & \textbf{37.92}  & 38.26  & 38.10 \\
    news  &   31.98  & 32.15  & 32.31  & \textbf{31.98} \\
    reading  &   40.56  & 40.43  & 40.16  & \textbf{39.01} \\
    talk  &   34.03  & \textbf{33.98}  & 34.29  & 34.14 \\
    variety  &   57.72  & \textbf{57.21}  & 58.18  & 57.29 \\
    others  &   34.65  & 33.60  & 33.97  & \textbf{33.47} \\
    \bottomrule
    \end{tabular}%
  \label{domain}%
\end{table}%

\begin{table*}[t]
  \centering
  \caption{ASR experiments on different decoding methods}
    \begin{tabular}{c|c|cccc}
    \hline
    \textbf{Dataset} & \textbf{Model} & \textbf{attention} & \textbf{CTC greedy search} & \textbf{CTC prefix beam search} & \textbf{attention rescoring} \\
    \hline
     & Baseline (CTC/AED)  & 4.86 & \textbf{4.82} & \textbf{4.82} & 4.56 \\
    Aishell-1 test & SememeASR-SP  & 4.87 & 4.91 & 4.91 & 4.53 \\
     & SememeASR-SE & 4.82 & 5.02 & 5.02 & 4.59\\
     & SememeASR-SEP & \textbf{4.71} & 4.94 & 4.94 & \textbf{4.53}\\
    \hline
    \hline
     & Baseline (CTC/AED)  & 12.53 & \textbf{12.67} & \textbf{12.67} & 12.03 \\
    Aishell-2 test & SememeASR-SP  & 12.57 & 12.79 & 12.79 & 12.16  \\
     & SememeASR-SE & 12.49 & 12.75 & 12.75 & \textbf{11.95} \\
     & SememeASR-SEP & \textbf{12.30} & 12.75 & 12.75 & 12.03 \\
    \hline
    \hline
     & Baseline (CTC/AED)  & 12.93 & 12.76 & 12.75 & 12.18 \\
    Aishell-2 dev & SememeASR-SP  & 12.43 & 12.85 & 12.84 & 11.99 \\
     & SememeASR-SE & 12.41 & \textbf{12.66} & \textbf{12.66} & 11.98 \\
     & SememeASR-SEP & \textbf{12.23} & 12.78 & 12.77 & \textbf{11.93} \\
    \hline
    \end{tabular}
  \label{different_decoding}%
\end{table*}%

To further illustrate the validity of our approach in more difficult text domains and more complex acoustic environments, we conduct further experiments on different domains of WenetSpeech.  Experimental  results in Table \ref{domain} indicate that our method also outperforms the baseline in new domains.

\subsection{Results of Long Tail Data}
To evaluate the ability of the model to recognize long-tail data, we first counted the long-tail data according to the methodology of \cite{deng2021alleviating}. The characters in the bottom 95\% of occurrences in the training set were used as long-tail characters. In addition, in order to analyze the impact of long-tail data in more detail, we have divided 10 intervals based on the ratio of the number of long-tail characters in the sentence. 
\begin{figure}[ht]   
	\centering
	\includegraphics[width=\linewidth,scale=1.00]{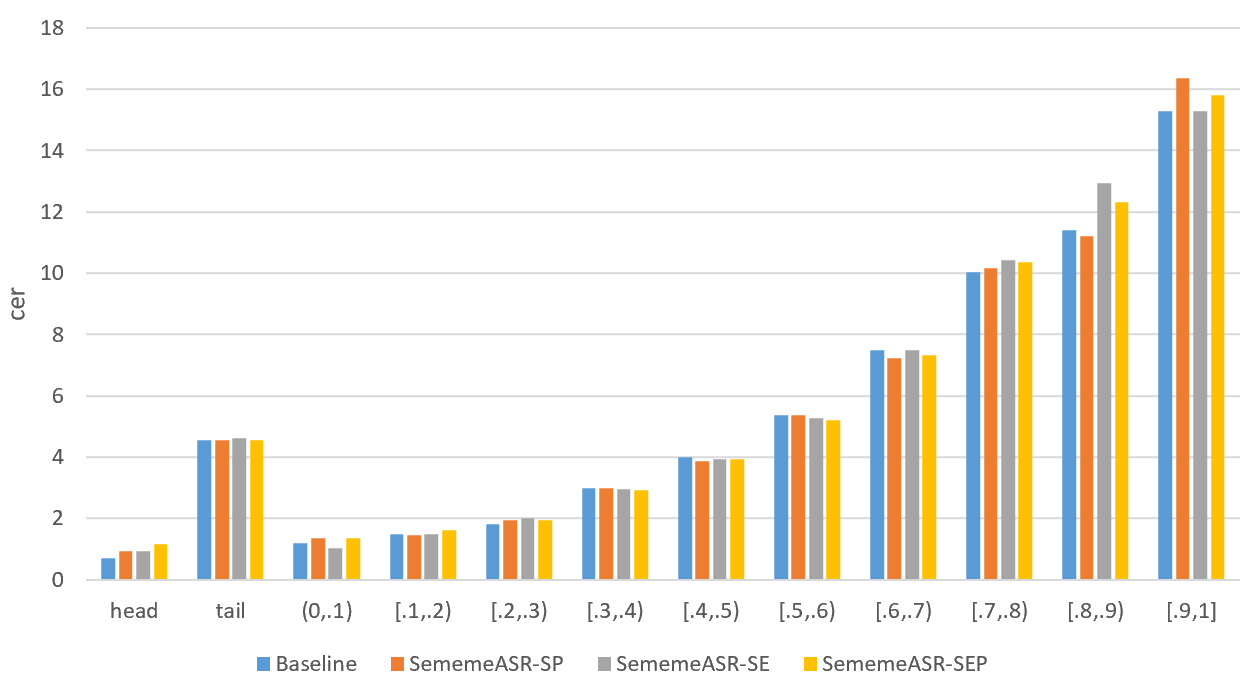}	
	\caption{CER of long-tailed data on Aishell-1 test set.}	
	\label{aishell-1 test set}	
 \vspace{-0.5cm}
\end{figure}

According to the result of Figure \ref{aishell-1 test set},  the proposed method can improve the recognition of long-tailed data, which will also lead to a little reduction in the recognition of head data.
However, as the proportion of long-tailed characters in a sentence increases, our model is less effective than the baseline model. It suggests that in the case of poor recognition, wrong results bring wrong semantic information, which hinders the role of semantic information.

\begin{figure}[h]   
	\centering
	\includegraphics[width=\linewidth,scale=1.00]{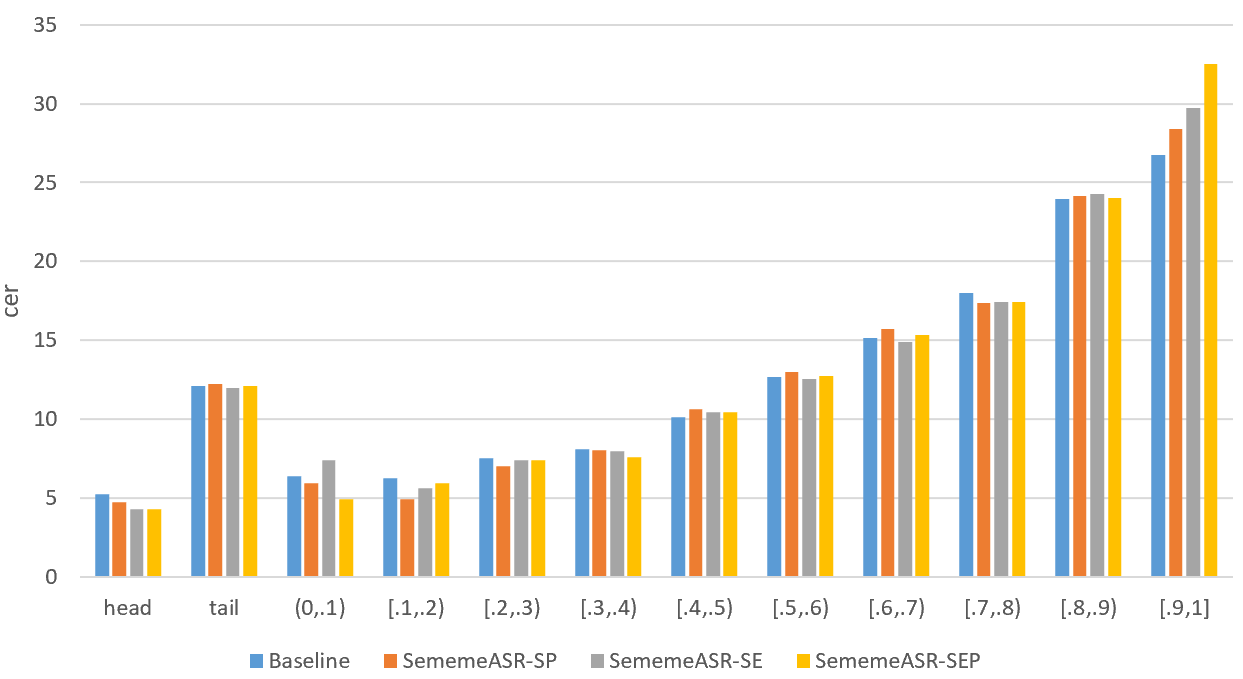}
	\caption{CER of long-tailed data on Aishell-2 test set.}	
	\label{aishell-2 test set}	
 \vspace{-0.5cm}
\end{figure}
Furthermore, we also evaluate the ability of the model in  recognizing long-tail data in the Aishell-2 dataset. As shown in Figure \ref{aishell-2 test set}, the proposed method can also improve the recognition of long-tail data overall on the new domain data. 

\subsection{Further Analysis on Different Parts of the Model}
To further analyze the effect of sememe information on the different parts of the model, we adopted different decoding approaches for our experiments.
Similar to 
\cite{yao21_interspeech}, we adopt four decoding methods, \textbf{attention}, \textbf{CTC greedy search}, \textbf{CTC prefix beam search}, \textbf{attention rescoring}. 

As shown in Table \ref{different_decoding}, in the \textbf{attention} and \textbf{attention rescoring} decoding methods, our proposed method performs better than the baseline CTC/AED model.
However, at the same time, the performance of the CTC-based decoding approach is inferior to that of the baseline. 
It indicates that our proposed approach increases the modeling capability of the \textit{Attention Decoder}, but has an impact on the \textit{Shared Encoder}. 
According to previous studies \cite{wang21t_interspeech, kim-etal-2022-joint}, this is because the \textit{Shared Encoder} part is more correlated with acoustic modeling, while the \textit{Attention Decoder} part is correlated with language modeling. 
Our approach improves the language modeling capability, but the coupling of acoustic modeling and language modeling makes our proposed method inevitably influential in its acoustic modeling component. And thus the effect of using the CTC decoding approach is degraded. 

\section{Conclusion and Future Work}
In this paper, we introduce sememe knowledge into the E2E ASR model and verify the effectiveness of external semantic knowledge for data-driven models. The proposed SememeASR can improve the recognition of long-tail data and enhance the domain generalization ability of the model.   
The intention of our work is to validate the effectiveness of sememe information for boosting ASR performance.
Therefore, we explore a series of  simple but effective model structures. In the future, we will consider optimizing the model structure and exploring different methods to further enhance the role of semantics for ASR.

\section{Acknowledgements}
This work is supported by Shenzhen Science and Technology Program (WDZC20200818121348001, JCYJ20220818101014 030), the Major Key Project of PCL (PCL2021A06, PCL2022 D01) and AMiner.Shenzhen SciBrain fund.


\bibliographystyle{IEEEtran}
\bibliography{mybib}

\end{document}